# Terahertz Magneto Optical Polarization Modulation Spectroscopy

D. K. George, A. V. Stier, C. T. Ellis,

B. D. McCombe, J. Černe and A. G. Markelz

*Department of Physics, University at Buffalo, The State University of New York, Buffalo, NY,*

*14260, USA*

*[*]Corresponding author: amarkelz@buffalo.edu*

We report the development of new terahertz techniques for rapidly measuring the complex Faraday angle in systems with broken time-reversal symmetry using the cyclotron resonance of a GaAs two-dimensional electron gas in a magnetic field as a system for demonstration of performance. We have made polarization modulation, high sensitivity (< 1 mrad) narrow band rotation measurements with a CW optically pumped molecular gas laser, and by combining the distinct advantages of terahertz (THz) time domain spectroscopy and polarization modulation techniques, we have demonstrated rapid broadband rotation measurements to < 5 mrad precision.

*OCIS codes:* 000.0000, 999.9999.

## Introduction

The DC conductivity tensor is the most direct measurement of the electronic response and is necessarily dependent on the underlying particle-particle correlations and interactions. The off diagonal conductivity elements $\sigma_{xy}$ can be determined by DC Hall measurements. However DC characterization provides a limited view of the interactions. The frequency dependence of the conductivity tensor is sensitive to single particle and collective excitations. Furthermore AC transport is less sensitive to sample imperfections and morphology, such as grain boundaries compared to DC transport. Of current interest is the nature and extent of localized states in 2D systems leading to quantum Hall plateaus in the terahertz range [1-3]. In addition, predictions [4] and measurements [5] of dramatic polarization changes in light reflected from topological insulators in the THz range motivate the development of high sensitivity measurements of the complete conductivity tensor. In this paper we introduce two techniques both based on polarization modulation: narrow band polarization modulated terahertz spectroscopy (NB-PMOTS) and broadband polarization modulated terahertz spectroscopy (BB-PMOTS). Few groups have successfully applied NB-PMOTS [6, 7] and BB-PMOTS with THz TDS has never before been demonstrated and provides unique possibilities for sample characterization.

The polarization of light changes when it interacts with an anisotropic system. In general, the Jones transmission matrix for a sample has complex elements $\tilde{t}_{ab} = t_{ab} e^{i\phi_{ab}}$ where $a$ and $b$ are the Cartesian coordinates $x$ and $y$, $t_{ab}$ is the real transmission amplitude, and $\phi_{ab}$ is the phase shift of the transmitted electric field. Using this formulation the transmitted electric field vector for linearly polarized incident light ($E_0$) is given by:

$$\vec{E_t} = \begin{pmatrix} \tilde{t}_{xx} & \tilde{t}_{xy} \\ \tilde{t}_{yx} & \tilde{t}_{yy} \end{pmatrix} \begin{pmatrix} E_o \\ 0 \end{pmatrix} = \begin{pmatrix} t_{xx}e^{i\phi_{xx}} & t_{xy}e^{i\phi_{xy}} \\ t_{yx}e^{i\phi_{yx}} & t_{yy}e^{i\phi_{yy}} \end{pmatrix} \begin{pmatrix} E_o \\ 0 \end{pmatrix}$$
$$= E_o \begin{pmatrix} t_{xx}e^{i\phi_{xx}} \\ t_{yx}e^{i\phi_{yx}} \end{pmatrix}. \qquad (1)$$

The transmitted polarization $\vec{E_t}$ differs from the original by the complex angle $\tilde{\theta}$ given in the small angle approximation by

$$\tan(\tilde{\theta}) = \frac{\tilde{t}_{xy}}{\tilde{t}_{xx}} \sim \tilde{\theta}. \qquad (2)$$

In general for small changes in polarization $\operatorname{Re}(\tilde{\theta})$ corresponds to rotation of the linear polarization, while $\operatorname{Im}(\tilde{\theta})$ is associated with the phase difference between the two orthogonal polarization components, which results in new ellipticity. This is shown in Fig.1a) where $\operatorname{Re}(\tilde{\theta})$ is the angle the major axis of the transmitted polarization ellipse makes with the initial polarization, whereas $\operatorname{Im}(\tilde{\theta})$ is the inverse tangent of the ratio of the minor to major axes. The picture is greatly simplified if the transmission coefficients are all real as seen in Fig.1b). The change in the polarization of the transmitted (reflected) light is given by the complex Faraday (Kerr) angle $\tilde{\theta}_F (\tilde{\theta}_K)$.

The application of a magnetic field perpendicular to a two dimensional free carrier system and parallel to the propagation direction of the probing radiation yields non-zero, complex, off-diagonal transmission matrix elements, and gives rise to the Faraday effect. For thin films the Faraday angle can be written in terms of the diagonal and off diagonal components of the magneto-optical conductivity tensor.

$$\tan(\tilde{\theta}_F) = \frac{\tilde{\sigma}_{xy}}{\tilde{\sigma}_{xx}} \left(1 + \frac{\tilde{n}_s + 1}{Z_0 \tilde{\sigma}_{xx} d}\right)^{-1}, \qquad (3)$$

as discussed by Kim et al. where $d$, $Z_0$ and $n_s$ are the film thickness, the impedance of free space, and the substrate index of refraction, respectively [8, 9]. Recalling the definition of the Hall angle $\tilde{\theta}_H$, where $\tan(\tilde{\theta}_H) = \dfrac{\tilde{\sigma}_{xy}}{\tilde{\sigma}_{xx}}$, one can see that the Faraday angle is the optical analog of the Hall angle.

Similarly the polarization of the light reflected from media immersed in an out-of-plane magnetic field has a net change in polarization defined by the complex Kerr angle $\tilde{\theta}_K$ which is related to the complex reflection coefficients and is proportional to the ratio of complex conductivities,

$$\tan(\tilde{\theta}_K) = \dfrac{\tilde{r}_{xy}}{\tilde{r}_{xx}} \propto \dfrac{\tilde{\sigma}_{xy}}{\tilde{\sigma}_{xx}^2}. \quad (4)$$

Measurement of both $\tilde{\theta}_F$ and $\tilde{\theta}_K$ allows one to determine the entire complex magneto-optical conductivity tensor without any additional measurements [8].

Strong magneto-optical effects in the terahertz range have been recently observed in graphene and topological insulators (TI). Rotation that exceeds $6°$ for a magnetic field of 7T has been reported in graphene [10]. In thin film TI, MacDonald et al. [4] predict that the Faraday rotation has a universal value $\text{Re}(\theta_F) = \tan^{-1}\alpha$, where $\alpha$ is the vacuum fine structure constant, and that TI can produce a giant Kerr rotation $\text{Re}(\theta_K) = \pi/2$. Aguilar et al. have reported rotation exceeding $65°$ for the TI $Bi_2Se_3$ [11]. Furthermore, predictions [3, 12] of interesting frequency dependent effects in the quantum Hall regime (remnants of quantum Hall plateaus) have been explored by measurements of $\tilde{\theta}_F$ in the THz range [2, 13, 14].

The two techniques discussed in this article are capable of determining $\tilde{\theta}$ by polarization modulation. The first employs a monochromatic source and measures the $\text{Re}\,\tilde{\theta}$, while the second is a new phase sensitive broadband technique that measures the full complex $\tilde{\theta}$.

## Narrow Band - Polarization Modulated Terahertz Spectroscopy (NB-PMOTS)

Narrow Band - Polarization Modulated Terahertz Spectroscopy (NB-PMOTS) is a monochromatic intensity measurement that is a starting point, both in terms of experimental technique and theoretical analysis, for BB-MOTS. NB-PMOTS directly determines the complex polarization angle $\theta$ given by Eq. (2). The real and imaginary parts of $\theta$ are determined by separate measurements that typically involve polarization modulation via spinning a linear polarizer and waveplate, respectively [15, 16]. Recently, Jenkins et al [6] have introduced a NB-PMOTS technique that allows simultaneous measurement of $\text{Re}\,\theta$ and $\text{Im}\,\theta$ by spinning a waveplate. For NB-PMOTS we focus on the determination of the real part of the $\theta$.

As depicted in Fig.2 NB-PMOTS consists of a linearly polarized, continuous wave monochromatic light source (MLS), a sample that affects the polarization via the complex transmission coefficients $\tilde{t}_{xx}$ and $\tilde{t}_{xy}$, a rotating polarizer, an AC coupled polarization insensitive detector (Det), and a lock-in amplifier referenced to twice the frequency of the rotating polarizer $(2\omega_{Rot})$. All polarization angles are measured relative to the initial vertical polarization of the linearly polarized light source.

To understand how changes in polarization are detected in this experimental configuration we employ a Jones calculus matrix train corresponding to the optical configuration. Beginning with an initially $\hat{x}$ polarized beam the electric field at the detector ($E_{det}$) is given by

$$E_{det} = \begin{pmatrix} 1 & 0 \\ 0 & 0 \end{pmatrix} \begin{pmatrix} \cos(\omega_{rot}t) & \sin(\omega_{rot}t) \\ -\sin(\omega_{rot}t) & \cos(\omega_{rot}t) \end{pmatrix} \begin{pmatrix} \tilde{t}_{xx} & \tilde{t}_{xy} \\ \tilde{t}_{yx} & \tilde{t}_{yy} \end{pmatrix} \begin{pmatrix} E_0 \\ 0 \end{pmatrix}. \quad (5)$$

Substituting the amplitude/phase form of the complex transmission coefficients into Eq. (5) yields a total intensity at the detector given by

$$\begin{aligned} I_{det} &\propto E_0^2 t_{xx}^2 \cos^2(\omega_{rot}t) \\ &+ 2E_0^2 t_{xx} t_{yx} \cos(\phi_{xx} - \phi_{xy}) \cos(\omega_{rot}t) \sin(\omega_{rot}t) \\ &+ E_0^2 t_{yx}^2 \sin^2(\omega_{rot}t) \\ &\propto \left( \frac{E_0^2 t_{xx}^2}{2} + \frac{E_0^2 t_{yx}^2}{2} \right) \\ &+ \left( \frac{E_0^2 t_{xx}^2}{2} - \frac{E_0^2 t_{yx}^2}{2} \right) \cos(2\omega_{rot}t) \\ &+ E_0^2 t_{xx} t_{yx} \cos(\phi_{xx} - \phi_{xy}) \sin(2\omega_{rot}t). \end{aligned} \quad (6)$$

Lock-in detection isolates the signal components at $2\omega_{rot}$, yielding the in-phase cosine and the out-of-phase sine components with amplitudes $S_{2\omega,X}$ and $S_{2\omega,Y}$, respectively:

$$S_{2\omega,X} = \frac{E_0^2}{2}\left(t_{xx}^2 - t_{yx}^2\right) \quad (7)$$

$$S_{2\omega,Y} = E_0^2 t_{yx} t_{xx} \cos(\phi_{xx} - \phi_{yx}). \quad (8)$$

For the case where $\tilde{t}_{xx}$ and $\tilde{t}_{yx}$ are real, no phase shift occurs (i.e. the sample adds no ellipticity). In this case the change in polarization $\tilde{\theta}$ becomes purely real as shown in Fig.1b). The transmission amplitudes reduce to $t_{xx}E_0 = E_t \cos(\theta)$ and $t_{yx}E_0 = E_t \sin(\theta)$. The tangent of the lock-in phase $\phi$ is given by $\dfrac{S_{2\omega,Y}}{S_{2\omega,X}}$, which is subsequently related to the polarization rotation angle by

$$\frac{S_{2\omega,Y}}{S_{2\omega,X}} = \tan(\phi) = \tan(2\theta). \qquad (9)$$

Thus for pure rotations the measured lock-in phase is exactly twice the polarization rotation. This can be understood pictorially via Fig. 3, where we compare the detector intensity versus time for two different polarizations. In Fig. 3a) the electric field incident on the rotator is vertically polarized and therefore the intensity at the detector is maximum when the rotating polarizer is vertical. In Fig. 3b) the sample rotates the initial polarization by $\theta$ away from vertical and now the maximum intensity at the detector occurs at a different orientation of the rotating polarizer and therefore at a different time, causing a phase shift in the resultant intensity trace.

In the most general case the sample transmission matrix is complex, yielding a complex change in polarization and Eq. (9) cannot be used. However for small rotations (i.e. $t_{xy} \ll t_{xx}$) the lock-in phase becomes

$$\begin{aligned}\tan(\phi) &= \frac{2 t_{xx} t_{yx}}{\left(t_{xx}^2 - t_{yx}^2\right)} \cos(\phi_{xx} - \phi_{xy}) \\ &\approx \frac{2 t_{yx}}{t_{xx}} \cos(\phi_{xx} - \phi_{yx}) = 2\,\mathrm{Re}\!\left(\frac{\tilde{t}_{yx}}{\tilde{t}_{xx}}\right) = 2\,\mathrm{Re}(\tilde{\theta}).\end{aligned} \qquad (10)$$

Thus for small changes in polarization and ellipticity the relationship of Eq. (9) is nearly recovered.

To demonstrate NB-PMOTS we use linearly polarized CW THz radiation from a $CO_2$ laser-pumped molecular gas laser (Edinburgh FIRL100). The light is focused by an off axis parabolic mirror onto the sample and then passes through the rotating polarizer to a liquid-helium cooled polarization insensitive Si bolometer (IR Labs). The rotating polarizer system consists of a polarizer mounted into a bearing housing that is rotated by a servo-motor (Smartmotor model

SM2316D-PLS2 by Animatics Corp). Unlike conventional AC motors, the phase noise, which is critical to these measurements, is very small (1 part in $10^4$) with the servo-motor. The rotation of the polarizer is monitored by an IR transceiver circuit that senses the reflected signal from the face of the rotating bearing. The bolometer signal is amplified by a specially designed low-noise preamplifier (with a cryogenic stage) and recorded through a Signal Recovery 7265 DSP lock-in amplifier.

We first demonstrate the technique using a static polarizer at the sample position. Fig.4a) shows the magnitude and half the phase of lock-in signal at $2\omega_{rot}$ as a function of time (recall Eq. (9)). When the static polarizer is rotated by 1°, the overall magnitude of the signal $\left(R = \sqrt{S_{2\omega,X}^2 + S_{2\omega,Y}^2}\right)$ remains essentially unchanged, however the lock-in phase changes proportionally with the rotation of the polarizer. This sensitivity of better than ± 1 mrad ~0.06° is achieved without any averaging other than the 200ms time constant of the lock-in. Although a higher rotation speed would reduce the 1/f noise, concerns over jitter in the delicate free standing wire grid polarizer as well as the motor/gear configuration limit the top speed.

We demonstrate the signal for a resonant system using the cyclotron resonance of a two dimensional electron gas (2DEG) as a function of magnetic field. In the semi-classical picture, a magnetic field perpendicular to the plane of a 2DEG leads carriers into circular orbits with a uniform helicity. In a 2DEG with parabolic bands in quantizing magnetic fields, the energy spacing between neighboring orbits (Landau levels) is constant and is given by $\hbar\omega_c$, where $\omega_c$ is the cyclotron frequency. Since these optical transitions are chiral, the circular symmetry about the magnetic field is broken and both the index of refraction and absorption are different for left- and right-circularly polarized light. The difference in index (circular birefringence) and the difference in absorption (circular dichroism) lead to the rotation and ellipticity, respectively, of the transmitted polarization as the magnetic field or photon energy is tuned through the cyclotron resonance. [17-19]. The sample is placed inside a 10T optical access superconducting magnet system (Oxford Spectromag). The radiation passes into the bore of the magnet through Kapton (outer) and c-cut sapphire (cold) windows. The radiation passes through the sample in the Faraday geometry. The resonant electronic system is a 2DEG formed at the interface of a GaAs/AlGaAs heterojunction about 55 nm below the surface by doping the $Al_{0.3}Ga_{0.7}As$ barrier with Si donors ($n_{2D}$ ($V_g$=0) = 5.7 x $10^{11}$ $cm^{-2}$ and 77K mobility of 177,000 $cm^2/Vs$). The sample is irradiated with monochromatic linearly polarized light at 2.52 THz. Figure 4b) shows the simultaneously recorded transmission and Faraday rotation as a function of magnetic field (for details see [14]). The transmittance shows resonant absorption due to the cyclotron resonance close to B=6.3 T, where a strong feature is observed in the Faraday rotation. A smaller absorption feature, barely visible in the transmittance, at about 3.9 T is clearly observed in the Faraday rotation signal. The 3.9 T feature is consistent with internal transitions in defect states [20-22]. Note that the signal to noise ratio for the 6.3 T (3.9 T) feature using transmission measurements is approximately 4:1 (1:1) whereas the signal to noise ratio for the same feature in Faraday rotation is more than 130:1 (40:1). This highlights the inherent advantage of the polarization sensitive technique as compared to conventional transmission measurements. A rotation of the linearly polarized light results from the breaking of time reversal symmetry in the sample, which produces a difference in left and right handed optical conductivities. Small changes in those quantities are therefore readily revealed in the latter technique since they are observed as a deviation from a null signal. Furthermore, since NB-PMOTS relies on the phase of the detector signal, it is not as sensitive to fluctuations in source intensity or detector sensitivity as transmission measurements.

## BB-PMOTS

Using a CW source allows one to achieve very high sensitivity at a single energy, however mapping out the energy dependence of the conductivity tensor using this method requires multiple measurements and changing/tuning sources. To rapidly measure spectral dependence one requires a broadband source and frequency sensitive detection, as can be achieved using terahertz time domain spectroscopy (THz TDS). In this work the source is a broadband THz pulse realized by current transient generation, and the THz pulse is mapped using electro-optic detection [23, 24]. In Fig. 5 we show a schematic of the THz TDS setup. A picosecond electric field pulse is generated by illuminating a photoconductive biased antenna fabricated on a semiconductor. The antenna has a gap, so that when unilluminated there is no current, but upon illumination with above bandgap light (typically near infrared, NIR), a fast current transient is generated, with a resultant radiated pulse. The pulse is short lived due to the short illuminating laser pulse (~ 100 fs) and the fast cancellation of the applied bias by the space charge field generated by the photogenerated electrons and holes. The electric field pulse is detected by co-propagating the THz pulse with a NIR probe pulse through an electro optic crystal. The NIR pulse probes the induced birefringence in the crystal by the time dependent THz electric field. The measured electric field pulse is shown in Fig. 5b). The spectral information is then determined by Fourier transform of the time domain pulse, with each frequency component $\nu$ having an amplitude and phase, see Fig. 5b).

Because THz TDS is a phase sensitive electric field measurement, the analysis is more straight-forward than NB-PMOTS, and one can access the full complex rotation angle in a single measurement with the rotating polarizer, eliminating the difficulties of calibrating and artifacts in using a rotating waveplate over a broad frequency range. The optical set up consists of a linearly polarized pulsed THz source, the rotating polarizer, the sample, a static linear polarizer and the phase sensitive electro optic detector. Jones matrix formulation for this optical is given by

$$E_{det} = \begin{pmatrix} 1 & 0 \\ 0 & 0 \end{pmatrix} \begin{pmatrix} \tilde{t}_{xx} & \tilde{t}_{xy} \\ \tilde{t}_{yx} & \tilde{t}_{yy} \end{pmatrix} \times \begin{pmatrix} cos\omega_{rot}t & sin\omega_{rot}t \\ -sin\omega_{rot}t & cos\omega_{rot}t \end{pmatrix}$$
$$\begin{pmatrix} 1 & 0 \\ 0 & 0 \end{pmatrix} \begin{pmatrix} cos\omega_{rot}t & -sin\omega_{rot}t \\ sin\omega_{rot}t & cos\omega_{rot}t \end{pmatrix} \begin{pmatrix} E_0 \\ 0 \end{pmatrix}.$$
(11)

We note that in this work, the linear polarizer before the detector is necessary due to the fact that electro optic detection is neither perfectly polarization sensitive nor insensitive [25]. Using a photoconductive switch antenna receiver, which is strongly polarization dependent, obviates the need for the final polarizer. However, the Jones matrix formalism above is the same for both cases. The electric field as detected by the electro-optic detector varies in time as

$$E_{det} = \tilde{t}_{xx}Ecos^2(\omega_{rot}t) + \tilde{t}_{xy}Esin(\omega_{rot}t)\,cos(\omega_{rot}t)$$
$$= \frac{\tilde{t}_{xx}E}{2} + \frac{\tilde{t}_{xx}E}{2}cos(2\omega_{rot}t) + \frac{\tilde{t}_{xy}E}{2}sin(2\omega_{rot}t).$$
(12)

In this case, one does not need to appeal to the small angle approximation to get a direct relationship to the Faraday angle. On Fourier transforming the time domain in-phase $cos(2\omega_{rot}t)$ and out-of-phase $sin(2\omega_{rot}t)$ waveforms and using the amplitude and phase form of the transmission coefficients we get

$$X(\nu) = t_{xx}(\nu)E_0(\nu)e^{i\phi_{xx}(\nu)}$$
$$Y(\nu) = t_{xy}(\nu)E_0(\nu)e^{i\phi_{xy}(\nu)}. \qquad (13)$$

Where $\tilde{t}_{xx}$ and $\tilde{t}_{yx}$ are the transmission coefficients of the sample. The ratio of the out of phase waveform to that of the in phase waveform yields

$$\frac{Y(\nu)}{X(\nu)} = \frac{t_{xy}(\nu)}{t_{xx}(\nu)} e^{i(\phi_{xy}(\nu) - \phi_{xx}(\nu))}. \qquad (14)$$

The real part of the above ratio yields the real Faraday rotation while the imaginary part yields the ellipticity.

For the measurements presented here, the THz TDS system used is shown in Fig.5a) and consists of a Ti-Sapphire laser (800nm, 100fs pulse width) beam focused between the electrodes of a photo conducting antenna (DC biased V = 40 V), which generates plane polarized THz pulses. The generated THz beam is steered by several off-axis parabolic mirrors through the rotating polarizer, the sample, and a final static polarizer before it is focused on to an electro optic detection setup. The final polarizer oriented along the polarization of the generated THz pulse makes sure that that there are no ambiguities in the polarization incident on the detector. The system has two foci. Situated at the first focus is a mechanical rotator with a free standing wire grid polarizer attached to its center. A servo motor connected to the rotator spins the polarizer up to a maximum speed of 4000 as was done with NB-PMOTS. The sample placed at the second focus inside a magneto optical cryostat capable of producing fields up to 10T. Again the sample is in the Faraday geometry.

We calibrated the system by measuring the rotations of a known polarizer, as was done with NB-PMOTS. A wire grid polarizer is placed at the sample position. Measurements were taken at polarizer orientations of 2°, 4° and 6° and the results were compared with the expected values. A systematic frequency-dependent background was determined and subtracted as a correction to the all the plots. The calibrated plots were accurate to within 5 mrad from 0.2 to 1.5 THz, as seen in Fig.6.

Next we measured a GaAs 2DEG sample at magnetic fields of ± 3.0T. To compare the results with a static polarizer method, where two consecutive measurements are done with two static wire grid polarizers using the technique first introduced by Spielman et al. [26] and described recently for the determination of $\tilde{\theta}_F$ and $\tilde{\theta}_K$ [2, 27]. Here the generating antenna is AC biased at 20 kHz and $V_{pp}$ = 120 V and the lock-in signal is now measured at $\omega_{bias}$. The rotator is replaced by a fixed polarizer at 45° from the output polarization of the generating antenna and a static polarizer is placed after the cryostat. This second polarizer is rotated either parallel or perpendicular to the first polarizer. Two separate waveform measurements are made: $E_{t\parallel}$ ($E_{t\perp}$) where the polarizers before and after the cryostat are parallel (perpendicular) to each other. The Faraday angle is now given by $\tan(\theta_F) = E_{t\perp} / E_{t\parallel}$. Both BB-PMOTS and static polarizer waveforms were taken with the same averaging (5 scans each). Figs. 7a) and 7b) gives the ellipticity and rotation due to the 2DEG. Solid curves represent rotator measurements while dashed curves correspond to static measurements. In comparison, the polarization modulation THz TDS gave slightly better sensitivity (<5mRad) as that of the static polarizer method when averaged similarly. Note that for the static polarizer measurement, two runs (parallel and perpendicular polarizers) are required to determine $\tilde{\theta}_F$ At the same time our technique has the advantage that the whole complex conductivity tensor is determined with a single scan avoiding any error introduced due to drift between the measurements.

**Summary and Conclusion**

We have demonstrated two techniques to measure the complex Faraday angle using polarization modulation. In the first narrow band technique we demonstrated a sensitivity of approximately 1 mrad in measuring the Faraday rotation while with the broad band polarization modulation THz TDS we were able to measure the Faraday rotation with an accuracy of $< 5$ mRad. We also compared our results with the static polarizer method. Unlike intensity measurements, the electric field sensitivity of the THz TDS allows crossed polarizers to be used to measure small $\theta_F$, but combining THz TDS with a rotating polarizer offers several additional advantages. First, BB-PMOTS has better signal to noise in half the data taking time and does not require manual realignment of polarizers for each run. This is particularly critical for the THz region, as these systems must be enclosed in purged environments making access to the optics more difficult. The time saving is especially important for low temperature and high magnetic field measurements due to the ever increasing cost of liquid helium. For crossed polarizers, the sensitivity of $\theta_F$ measurements is limited by leakage through the polarizers whereas in measurements using a rotating polarizer leakage only reduces the overall amplitude of the modulation but does not affect the phase of the signal, which determines $\theta$. We note that BB-MOTS uses a DC THz generating antenna bias and the modulation frequency is limited by the servo motor, (typically $< 150$ Hz), whereas the modulation for static polarizer measurements comes from the AC bias of the antenna (typically 20-100 kHz). While the static polarizer technique can take advantage of this high modulation frequency to sensitively measure the transmitted electric field, $\theta$ in this case is determined by two measurements with different polarizer configurations that are separated by at least several minutes. On the other hand, the rotator modulates the polarization configuration at 150 Hz and therefore is much less susceptible to drift. The demonstrated sensitivity of NB-PMOTS and BB-PMOTS was achieved with minimum averaging, and the sensitivity can readily be improved with increased averaging. Currently determination of the complex off diagonal conductivity in the 0.2 – 1.5 THz range is limited by instrumental challenges. This work adds a new capability to this interesting frequency range, and will considerably improve our ability to characterize new materials.

# REFERENCES

1. T. Morimoto, Y. Hatsugai, and H. Aoki, "Optical Hall Conductivity in Ordinary and Graphene Quantum Hall Systems " Physical Review Letters **103**, 116803 (2009).
2. Y. Ikebe, T. Morimoto, R. Masutomi, T. Okamoto, H. Aoki, and R. Shimano, "Optical Hall Effect in the Integer Quantum Hall Regime," Physical Review Letters **104**, 256802 (2010).
3. T. Morimoto, Y. Avishai, and H. Aoki, "Dynamical scaling analysis of the optical Hall conductivity in the quantum Hall regime," Physical Review B **82**, 081404R (2010).
4. W.-K. Tse and A. H. MacDonald, "Giant Magneto-Optical Kerr Effect and Universal Faraday Effect in Thin-Film Topological Insulators," Physical Review Letters **105**, 057401 (2010).
5. R. V. Aguilar, A. V. Stier, W. Liu, L. S. Bilbro, D. K. George, N. Bansal, L. Wu, J. Cerne, A. G. Markelz, S. Oh, and N. P. Armitage, "THz response and colossal Kerr rotation from the surface states of the topological insulator $Bi_2Se_3$," arXiv:1105.0237, Submitted to Phys. Rev. Lett. 2010 (2010).
6. G. S. Jenkins, D. C. Schmadel, and H. D. Drew, "Simultaneous measurement of circular dichroism and Faraday rotation at terahertz frequencies utilizing electric field sensitive detection via polarization," Rev. Sci. Instrum. **81**(2010).
7. D. C. Schmadel, G. S. Jenkins, J. J. Tu, G. D. Gu, H. Kontani, and H. D. Drew, "Infrared Hall conductivity in optimally doped $Bi_{2}Sr_{2}CaCu_{2}O_{8+\delta}$: Drude behavior examined by experiment and fluctuation-exchange-model calculations," Physical Review B **75**, 140506 (2007).
8. M.-H. Kim, G. Acbas, M.-H. Yang, I. Ohkubo, H. Christen, D. Mandrus, M. A. Scarpulla, O. D. Dubon, Z. Schlesinger, P. Khalifah, and J. Cerne, "Determination of the infrared complex magnetoconductivity tensor in itinerant ferromagnets from Faraday and Kerr measurements," Physical Review B **75**, 214416 (2007).
9. J. Černe, D. C. Schmadel, M. Grayson, G. S. Jenkins, J. R. Simpson, and H. D. Drew, "Midinfrared Hall effect in thin-film metals: Probing the Fermi surface anisotropy in Au and Cu," Phys. Rev. B **61**, 8133 (2000).
10. I. Crassee, J. Levallois, A. L. Walter, M. Ostler, A. Bostwick, E. Rotenberg, T. Seyller, D. van der Marel, and A. B. Kuzmenko, "Giant Faraday rotation in single- and multilayer graphene," Nat Phys **7**, 48-51 (2011).
11. R. V. Aguilar, A. V. Stier, W. Liu, L. S. Bilbro, D. K. George, N. Bansal, L. Wu, J. Cerne, A. G. Markelz, S. Oh, and N. P. Armitage, "THz response and colossal Kerr rotation from the surface states of the topological insulator $Bi_2Se_3$," arXiv:1105.0237 (2010).
12. T. Morimoto, Y. Hatsugai, and H. Aoki, "Optical Hall Conductivity in Ordinary and Graphene Quantum Hall Systems " Physical Review Letters, vol. 103, p. 116803, 2009. **103**, 116803 (2009).
13. A. V. Stier, H. Zhang, C. T. Ellis, D. Eason, G. Strasser, B. D. McCombe, and J. Cerne, "THz Quantum Hall Conductivity Studies in a GaAs Heterojunction," AIP Conference Proceedings **1399**, 627-628 (2011).
14. A. V. Stier, H. Zhang, C. T. Ellis, D. Eason, G. Strasser, B. D. McCombe, T. Morimoto, H. Aoki, and J. Cerne, "Quantum effects in the terahertz optical Hall conductivity at the cyclotron resonance of a two dimensional electron gas," arXiv:1201.0182, Submitted to Phys. Rev. Lett. 2012 (2012).
15. M. Grayson, L. B. Rigal, D. C. Schmadel, H. D. Drew, and P.-J. Kung, "Spectral Measurement of the Hall Angle Response in Normal State Cuprate Superconductors," Physical Review Letters **89**, 037003 (2002).
16. G. S. Jenkins, D. C. Schmadel, and H. D. Drew, "Simultaneous measurement of circular dichroism and Faraday rotation at terahertz frequencies utilizing electric field sensitive detection via polarization," Rev. Sci. Instrum. **81**, 083903 (2010).
17. K. W. Chiu, T. K. Lee, and J. J. Quinn, "Infrared magneto-transmittance of a two-dimensional electron gas," Surface Science **58**, 182-184 (1976).
18. H. Piller, "Effect of Internal Reflection on Optical Faraday Rotation," Journal of Applied Physics **37**, 763-767 (1966).
19. H. Piller, "Far infrared Faraday rotation in a two-dimensional electron gas," Journal of Vacuum Science and Technology **16**, 2096-2100 (1979).
20. J. P. Cheng, Y. J. Wang, B. D. McCombe, and W. Schaff, "Many-electron effects on quasi-two-dimensional shallow-donor impurity states in high magnetic fields," Physical Review Letters **70**, 489-492 (1993).
21. S. Huant, S. P. Najda, and B. Etienne, "Two-dimensional D^{-} centers," Physical Review Letters **65**, 1486-1489 (1990).
22. J. Kono, S. T. Lee, M. S. Salib, G. S. Herold, A. Petrou, and B. D. McCombe, "Optically detected far-infrared resonances in doped GaAs quantum wells," Physical Review B **52**, R8654-R8657 (1995).
23. D. Grischkowsky, S. Keiding, M. VanExter, and C. Fattinger, "Far-infrared time-domain spectroscopy with terahertz beams of dielectrics and semiconductors," Journal of the Optical Society of America B: Optical Physics **7**, 2006-2015 (1990).
24. Q. Wu and X.-C. Zhang, "Free-space electro-optic sampling of terahertz beams," Appl. Phys. Lett. **67**, 3523-3525 (1995).
25. P. C. M. Planken, H.-K. Nienhuys, H. J. Bakker, and T. Wenckebach, "Measurement and calculation of the orientation dependence of terahertz pulse detection in ZnTe," J. Opt. Soc. Am. B **18**, 313-317 (2001).
26. S. Spielman, B. Parks, J. Orenstein, D. T. Nemeth, F. Ludwig, J. Clarke, P. Merchant, and D. J. Lew, "Observation of the Quasi-Particle Hall-Effect In Superconducting YBa2Cu3O7-δ," Physical Review Letters **73**, 1537-1540 (1994).
27. K. Yatsugi, N. Matsumoto, T. Nagashima, and M. Hangyo, "Transport properties of free carriers in semiconductors studied by terahertz time-domain magneto-optical ellipsometry," Appl. Phys. Lett. **98**, 212108 (2011).

# CAPTIONS

Figure 1.
Schematics of the polarization change for a sample with off-diagonal transmission coefficients. a) Shows that for complex $\tilde{t}_{ab} = t_{ab} e^{i\phi_{ab}}$, $\boldsymbol{E}_t$ has a component perpendicular to the incident polarization and this $E_{ty}=E_o t_{yx}$ is shifted by a distance $\Delta L$, which produces a phase delay relative to $E_{tx}$. $\mathrm{Re}(\theta)$ is the rotation and $\mathrm{Im}(\theta)$ is the ellipticity of the transmitted polarization. b) shows when all $t_{ab}$ are real there is a net magnitude change and a pure rotation.

Figure 2.
Schematic of the NB-PMOTS.

Figure 1
The intensity at the detector, $I_{det}$, as a function of time is shown for two different incident polarization orientations. $P_{rot}$ indicates the orientation of the rotating polarizer. For a) the incident polarization is vertical corresponding to the t = 0 orientation of the rotating polarizer and the phase of the $2\omega_{rot}$ signal $\phi = 0$. For b) the incident polarization is rotated from vertical by $\theta$ and the phase of the $2\omega_{rot}$ signal $\phi = 2\theta$, see Eq. (9).

Figure 2.
Data for NB-PMOTS. a) Calibration of the detected polarization rotation using a static polarizer at fixed angles. Note there is no noticeable transmission change for the 1° rotation occurring at 0.1 min., however the Faraday rotation determined from the lock-in phase clearly resolves this rotation to ~ 0.05 ° sensitivity. b) NB-PMOTS for a GaAs 2DEG as a function of magnetic field. Both a low field defect state (3.9 T) and the cyclotron resonance (6.3 T) are observed. Note that while the 3.9 T feature is barely resolved in the transmittance, it is clearly shown in the Faraday rotation.

Figure 5.
a) The experimental setup for BB-PMOTS and b) Extraction of frequency spectrum from THz pulse

Figure 6.
Measured rotation spectra using BB-PMOTS for three static polarizer orientations. The label on the curve is the actual rotation of the static polarizer.

Figure 7.
Rotation a) and ellipticity b) near the cyclotron frequency for 2DEG in GaAs at ± 3Tesla. The solid lines are for BB-PMOTS and the dashed lines are for static crossed polarizer measurements.

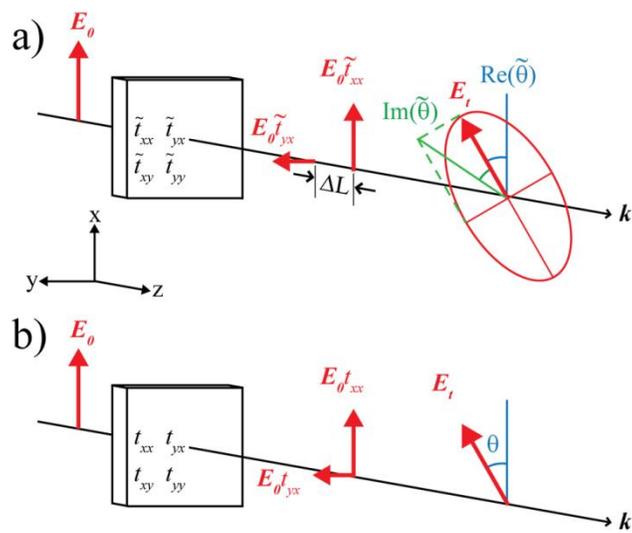

Figure 1

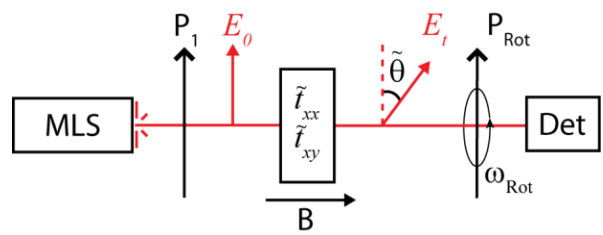

Figure 2

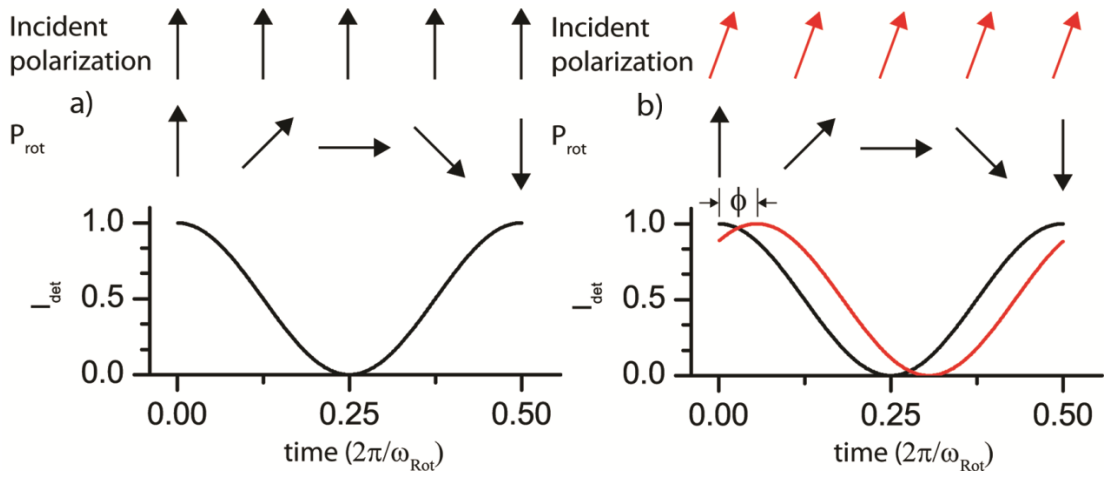

Figure 3

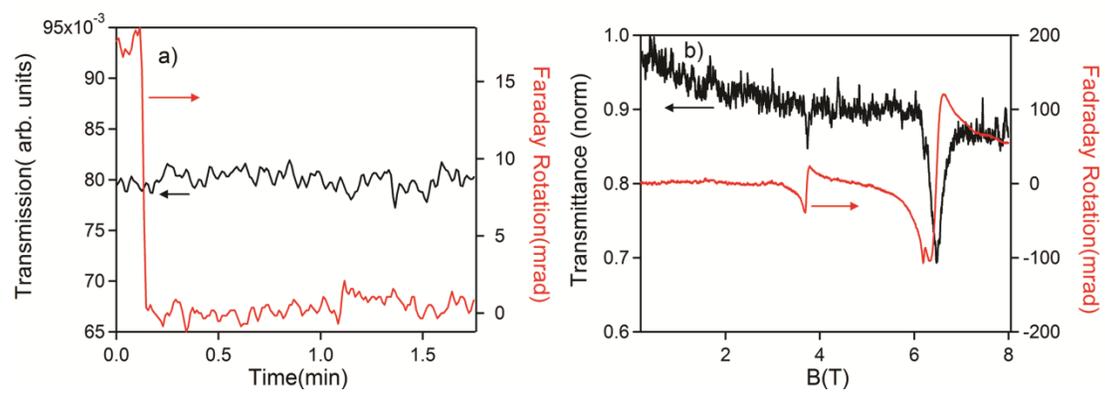

Figure 4

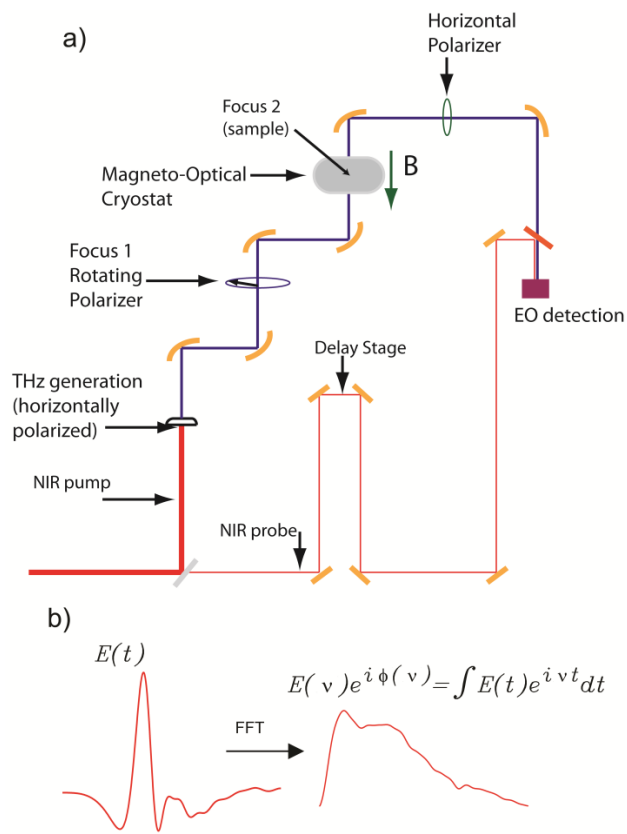

Figure 5

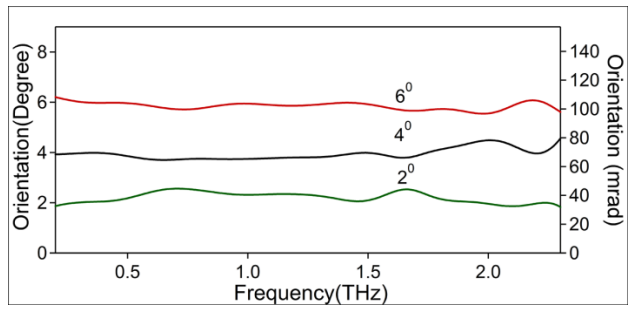

Figure 6

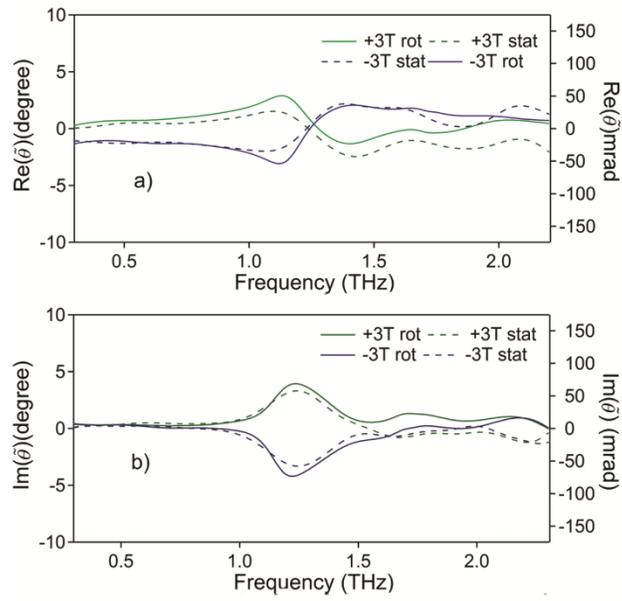

Figure 7